\documentstyle[12pt,epsf]{article}
\newcommand{\beq}{\begin{equation}}
\newcommand{\eeq}{\end{equation}}
\def\bea{\begin{eqnarray}}
\def\eea{\end{eqnarray}}
\newcommand{\ba}{\begin{array}}
\newcommand{\ea}{\end{array}}
\newcommand{\nn}{\nonumber}
\newcommand{\mathbold}{\bf}

\def\et{{\it et al.}}

\begin{document}

\title{\vskip-2.5truecm{\hfill \baselineskip 14pt {{
\small  \\    \hfill FTUV/98-23\\
\hfill IFIC/98-23 \\
\hfill MZ-TH/98-01 \\ 
\hfill May 1998}}\vskip .9truecm}
 {\bf Electroweak Phase Transition in Left-Right 
Symmetric Models}}

\vspace{5cm}

\author{Gabriela Barenboim$^1$ and Nuria Rius$^2$
 \\  \  \\
{\it 1) Institut fur Physik - Theoretische 
Elementarteilchenphysik }\\
{\it Johannes Gutenberg Universitat, D-55099 Mainz, Germany}
\\
\\
{\it 2) Depto. de F\'\i sica Te\'orica and IFIC, Centro Mixto }\\
{\it Universidad de Valencia-CSIC, Valencia, Spain}\\
}

\date{}
\maketitle
\vfill

\begin{abstract}
\baselineskip 20pt
We study the finite-temperature effective potential of minimal
left-right symmetric models containing a bidoublet and two triplets
in the scalar sector. 
We perform a numerical analysis of the parameter space compatible 
with the requirement that baryon asymmetry is not washed out by 
sphaleron processes after the electroweak phase transition. 
We find that the spectrum of scalar particles for these acceptable
cases is consistent with present experimental bounds.
\end{abstract}
\vfill
\thispagestyle{empty}

\newpage
\pagestyle{plain}
\setcounter{page}{1}

\renewcommand{\thefootnote}{\arabic{footnote}}
\setcounter{footnote}{0}

\section{Introduction}

The origin of the observed baryon asymmetry of the Universe (BAU)
remains an interesting open question in particle physics. 
In 1967, Sakharov \cite{sak} established the three basic requirements 
for obtaining this baryon asymmetry as a result of particle 
interactions in the early universe:  
a) Baryon number violation,
b) C and CP violation, 
c) departure from thermal equilibrium.
These conditions are fulfilled in grand unification theories, 
in which the baryon asymmetry is generated by the out-of-equilibrium 
B-violating decay of some superheavy boson. However, this 
scenario presents the problem that anomalous processes can partially 
or totally erase the baryon asymmetry generated at extremely 
high energies. 

It was realized in \cite{krs} that Sakharov conditions may also 
be satisfied at weak scale temperatures, if the
electroweak phase transition is first order.
In a strongly first order electroweak phase transition, 
bubbles of the true ground state (broken phase) nucleate 
and expand until they fill the Universe; local departure from 
thermal equilibrium occurs in the vicinity of the expanding 
bubble walls. C and CP are known to be violated by the electroweak 
interactions, and anomalous baryon number violation is fast at high 
temperatures in the symmetric phase.
Moreover, electroweak baryogenesis provides an explanation of 
the observed BAU in terms of experimentally accessible physics and 
hence much attention has been devoted to the study of this
possibility \cite{ckn}.

In principle, the Standard Model (SM) contains all the necessary 
ingredients for electroweak baryogenesis, but it has two problems:
the CP asymmetry induced by  the Kobayashi-Maskawa phase is 
far too small to account for the observed $n_B / s$ ratio \cite{b2}, 
and the phase transition appears too weakly first order for the 
Higgs mass experimentally allowed \cite{pt}. 
To avoid the erasure of the baryon asymmetry produced during 
the phase transition, the sphaleron processes need to be
sufficiently suppressed in the broken phase 
and this in turn is directly related to the strength of the 
phase transition. 
Quantitatively, the requirement is that the ratio of the vacuum 
expectation value of the Higgs field at the critical temperature 
to the critical temperature must be larger than one, 
\beq
\frac {v(T_c)}{T_c} > 1  \ .
\label{sph}
\eeq
In the SM, this imposes an upper bound on the Higgs mass which 
is below the present experimental bound  
$m_H \ge 88$ GeV \cite{al}.

However these two problems may be absent in  
simple extensions of the SM, which contain additional 
sources of CP violation and more scalars than the SM.
The larger parameter space in the scalar sector allows
for a stronger first-order phase transition without such a light Higgs
\cite{ah}. 
Several possibilities have been analyzed in detail: 
two Higgs models with a strong CP phase \cite{nkc}-\cite{cpr}, 
heavy neutrinos \cite{ng}, and supersymmetric models \cite{susy}.

In the present paper, we consider one of the most attractive 
extensions of the SM, based on the gauge group 
$SU(2)_L \times SU(2)_R \times U(1)_{B-L}$ \cite{ps,sm}. 
Various different models employing this gauge group are possible, 
depending on which Higgs and fermion spectrum is chosen, and on 
whether or not exact discrete left-right symmetry is imposed. 
We are interested in the class of left-right symmetric models 
described in \cite{ps,ms}.  
Besides the original idea of explaining the observed parity violation 
of the weak interaction at low energies, these models provide also an 
explanation for the lightness of the ordinary neutrinos, 
via the so-called see-saw mechanism.

There are two possible scenarios for baryogenesis in left-right 
symmetric models. In the first one, the BAU is generated at the 
scale where local $B-L$ is broken. This in turn can occur 
during the right phase transition (if it is first-order), 
due to the reflection of 
right-handed neutrinos on walls of broken phase bubbles 
(i.e. $SU(2)_L \otimes U(1)_Y$ symmetric) \cite{jm}, 
and/or via out-of-equilibrium decay of right handed Majorana 
neutrinos \cite{mz}.
In both cases, a lepton number asymmetry is produced and  
subsequently converted into a baryon number one  
through the rapid ($B+L$)-violating anomalous processes 
above the electroweak phase transition temperature.  
However, the right-handed scale needed to account for the BAU
(${\cal O} (10$ TeV) in the first case and above $10^6 - 10^7$ GeV in 
the latter) is too high to have any low-energy observable implication.
In the second scenario, baryogenesis takes place at the lower 
electroweak scale, mainly due to the reflection of the top quark
on walls of the true vacuum (i.e. $U(1)_{em}$ symmetric) bubbles.
There are estimates \cite{mz,jm,rt} of the baryon asymmetry produced 
in left-right symmetric models using this mechanism, 
although they neglect effects that are now understood to be
important, such as diffusion \cite{jpt}, \cite{diffu} 
and thermal scattering \cite{b2}.

We focus on the possibility of electroweak baryogenesis.
We perform an analysis of the electroweak phase transition 
in phenomenologically acceptable left-right symmetric models
with a relatively low right scale, 
which have interesting implications in present and planned 
experiments \cite{phen}
\footnote {A similar study in a left-right symmetric model with a 
simpler scalar content has been done in ref. \cite{cv}.}.

It has been shown recently that, contrary to previous belief, 
spontaneous CP violation can occur in the minimal left-right 
symmetric model considered here \cite{bb}.
However baryogenesis with (only) spontaneous breakdown of 
CP presents severe cosmological problems, due to the formation
of domain walls as a result of the breaking of a discrete 
symmetry. Although this problem can be solved, in order to generate 
the BAU the scale of spontaneous CP violation and the scale at which 
baryogenesis takes place must be different \cite{rt}; otherwise, 
an equal amount of matter and anti-matter is generated. 
In the minimal left-right symmetric model
with spontaneous CP-violation both scales coincide and
therefore electroweak baryogenesis is not feasible.

The remainder of this paper is structured as follows. 
In section 2 we describe the model, while the effective potential 
at finite temperature is calculated in section 3. 
The order of the electroweak phase transition is analyzed in 
section 4 and we conclude in section 5.

\section{Left-right symmetric model}

We consider the minimal 
$SU(2)_L \times SU(2)_R \times U(1)_{B-L}$ model with a left-right 
discrete symmetry \cite{ms,despa}.
This model is formulated so that parity is a spontaneously  
broken symmetry: the Lagrangian is left-right symmetric but the
vacuum is not invariant under the parity transformation.
Thus, the observed V-A structure of the weak interactions 
is only a low energy phenomenon, which should disappear when 
one reaches energies of order $v_R$, where $v_R$ is the 
vacuum expectation value of some right-handed scalar.

According to the left-right symmetry requirements, quarks (and
similarly leptons) are placed in left and right doublets,
\bea
\Psi_{iL} =\pmatrix{u_i\cr d_i}_L \equiv \left( 2, 1,\frac{1}{3} \right),
\;\;\;
\Psi_{iR} =\pmatrix{u_i\cr d_i}_R \equiv \left(1, 2,\frac{1}{3} \right),
\eea
where $i=1,2,3$ is the generation index and the representation content
with respect to the gauge group is explicitly given.
Taking advantage of the fact that the weak interactions
observed at low energies involve only the left handed helicity components, 
the electric charge formula can also be written in a left-right symmetric form
as
\bea
Q= I_{3L}^W + I_{3R}^W + \frac{B-L}{2}
\label{elc}
\eea
where $ I_{3}^W$ denotes the third component of the weak isospin.
 
Regarding the bosons, 
gauge vector bosons consist of two triplets ${\mathbold  W}_L^\mu
 \equiv (3,1,0)$,
${\mathbold  W}_R^\mu \equiv (1,3,0)$, and a singlet $B^\mu \equiv (1,1,0)$.

The Higgs sector of the model is dictated by two requirements,the 
choice of the symmetry breaking chain and the desire to reproduce the 
phenomenologically observed light masses of the known neutrinos via the 
so-called see-saw mechanism.   The best candidates for these purposes
seem to be
\bea
\Phi &=&
\left(
\ba{cc}
\phi_1^0 & \phi_1^+  \\ 
\phi_2^- & \phi_2^0
\ea \right)\equiv \left(\frac 1 2, \frac 1 2^* , 0 \right) 
\\
\Delta_L &=& \pmatrix{ \frac{\delta_L^+}{\sqrt{2}} & \delta_L^{++}\cr
\delta_L^0 &  -\frac{\delta_L^+}{\sqrt{2}}}  \equiv    (1,0,2)  
\\
\Delta_R &=& \pmatrix{\frac{\delta_R^+}{\sqrt{2}} & \delta_R^{++}\cr
\delta_R^0 &  -\frac{\delta_R^+}{\sqrt{2}}}  \equiv   (0,1,2)
\eea
where the scalar fields have been written in a convenient
representation given by $2\times2$ matrices.

Let us now discuss the form of the Lagrangian. 
We require the Lagrangian 
to be invariant under the discrete left-right symmetry defined by
\beq
\Psi_L \leftrightarrow  \Psi_R  \;\;\;\;\;
\Delta_L \leftrightarrow \Delta_R   \;\;\;\;\;
\Phi \leftrightarrow \Phi^{\dagger}
\eeq
where $\Psi$ denotes any fermion. We assume that the global phases 
allowed to appear in the transformations above are absorbed by proper
redefinition of the fields.

The most general renormalizable Lagrangian consistent with the 
above discrete symmetry and gauge invariance can be written as 
\beq
{\cal L} = {\cal L}_{gauge} + {\cal L}_{f} + {\cal L}_{Higgs} 
\label{lag}
\eeq
where the gauge field part of the Lagrangian contains the kinetic
energy terms for the gauge bosons corresponding to the gauge groups
$SU(2)_L \times SU(2)_R \times U(1)_{B-L}$. The gauge coupling 
constants for the gauge groups $SU(2)_L$ and $SU(2)_R$ are the same 
and we denote it by $g$, while that of the $U(1)_{B-L}$ is denoted by 
$g^\prime $. The fermionic part of the Lagrangian,
${\cal L}_{f}$, contains the kinetic
energy terms for the fermions and the Yukawa couplings,
which are given by 
\bea
-{\cal L}_Y= \sum_{a,b} h_{ab} \overline{\Psi}_{aL} \Phi \Psi_{bR} +
\tilde{h}_{ab} \overline{\Psi}_{aL} \tilde{\Phi} \Psi_{bR} + 
i f_{ab} \left[ \Psi^T_{aL} C \tau_2 \Delta_L
  \Psi_{bL} +
(L \leftrightarrow R )\right] + \mbox{h.c.}
\label{y}
\eea
where $\tilde{\Phi} = \tau_2 \Phi^* \tau_2$, $C$ is the Dirac 
charge-conjugation matrix and $a,b$ label different generations. 

%However, the neutrino sector
%will not be relevant for our analysis (the masses are either too high 
%(right-handed neutrinos) 
%or too low (left-handed ones)
% to become relevant) and therefore only $h$ and $\tilde{h}$ 
%will be the parameters that will enter in our calculations.
 
The Higgs part
of the Lagrangian contains the kinetic energy terms for the fields
$\Delta_{L,R}$ and $\Phi$ and the scalar interaction terms, i.e. the most
general scalar potential. 
This potential cannot 
have trilinear terms: because the nonzero $B-L$
quantum numbers of the $\Delta_L$ and
$\Delta_R$ triplets, these must always appear in the quadratic combinations 
$\Delta_L^\dagger \Delta_L $, $\Delta_R^\dagger \Delta_R $,
$\Delta_L^\dagger \Delta_R $ or $\Delta_R^\dagger \Delta_L $.
These can never be combined with a single bidoublet $\Phi$ 
in such a way as to form  $SU(2)_L$ and $SU(2)_R$ singlets.
Nor can three bidoublets be combined so as to yield a singlet.
However, quartic combinations of the form $\mbox{Tr}( \Delta_L^\dagger \Phi
\Delta_R \Phi^\dagger )$  are in general allowed by the left-right symmetry.
According to  these strict conditions, the most general form of the Higgs 
potential is
\beq
{\mathbold V } = {\mathbold V}_\Phi  + {\mathbold V}_\Delta  +
{\mathbold V}_{\Phi \Delta}  
\label{pot}
\eeq
with 
\bea
{\mathbold V}_\Phi & = & -\mu_1^2 \, \mbox{Tr} (\Phi^\dagger \Phi ) \, 
- \, \mu_2^2 \left[ \mbox{Tr} (\tilde{\Phi} \Phi^\dagger ) + 
 \mbox{Tr}(\tilde{\Phi}^\dagger \Phi ) \right] \, + \, 
 \lambda_1 \, \left[ \mbox{Tr}(\Phi \Phi^\dagger ) \right]^2 \, 
 + \,  \nn \cr
 && \lambda_2 \, \left\{ \left[ \mbox{Tr}(\tilde{\Phi} 
 \Phi^\dagger ) \right]^2 +
 \left[ \mbox{Tr}(\tilde{\Phi}^\dagger \Phi ) \right]^2 \right\} \, + \, 
 \lambda_3 \, \left[ \mbox{Tr}(\tilde{\Phi} \Phi^\dagger )  
 \mbox{Tr}(\tilde{\Phi}^\dagger  \Phi)\right]  \, + \, \nn   \\ 
  & & \lambda_4 \, \left\{ \mbox{Tr}(\Phi^\dagger \Phi )
\left[ \mbox{Tr}(\tilde{\Phi} \Phi^\dagger ) +
 \mbox{Tr}(\tilde{\Phi}^\dagger \Phi ) \right]  \right\}
\nn \eea

\bea
{\mathbold V}_\Delta  & = & -\mu_3^2 \left[ 
\mbox{Tr}(\Delta_L \Delta_L^\dagger ) +
 \mbox{Tr}(\Delta_R \Delta_R^\dagger ) \right] \, + \,
 \rho_1 \,  \left\{ \left[ 
 \mbox{Tr}(\Delta_L \Delta_L^\dagger ) \right]^2  + \right. \nn \\
&&  \left. \left[ \mbox{Tr}(\Delta_R 
 \Delta_R^\dagger ) \right]^2 \right\} \, + \,   
\rho_2 \, \left[ \mbox{Tr}(\Delta_L \Delta_L ) 
 \mbox{Tr}(\Delta_L^\dagger \Delta_L^\dagger ) +  \right. \nn \\
&& \left. \mbox{Tr}(\Delta_R \Delta_R ) 
 \mbox{Tr}(\Delta_R^\dagger \Delta_R^\dagger ) \right] \, + \, 
 \rho_3 \, \left[ \mbox{Tr}(\Delta_L \Delta_L^\dagger ) 
 \mbox{Tr}(\Delta_R \Delta_R^\dagger ) \right] \, + \, \nn  \\  
   & &  \rho_4 \, \left[ \mbox{Tr}(\Delta_L \Delta_L ) 
 \mbox{Tr}(\Delta_R^\dagger \Delta_R^\dagger ) +  
 \mbox{Tr}(\Delta_L^\dagger \Delta_L^\dagger ) 
 \mbox{Tr}(\Delta_R \Delta_R ) 
  \right] 
\nn \eea

\bea
{\mathbold V}_{\Phi \Delta} & = &
\alpha_1 \, \left\{ \mbox{Tr}(\Phi^\dagger \Phi )
\left[ \mbox{Tr}(\Delta_L \Delta_L^\dagger ) +
 \mbox{Tr}(\Delta_R \Delta_R^\dagger ) \right]  \right\} \, + \,
\alpha_2 \, \left[ \mbox{Tr}(\tilde{\Phi}^\dagger \Phi ) 
\mbox{Tr}(\Delta_R \Delta_R^\dagger ) \right.  
\nn \\
&&  \left.   
+    \mbox{Tr}(\tilde{\Phi} \Phi^\dagger ) 
 \mbox{Tr}(\Delta_L \Delta_L^\dagger ) \right] 
 + \alpha_2^* \left[  
  \mbox{Tr}(\tilde{\Phi} \Phi^\dagger ) \mbox{Tr}
  (\Delta_R \Delta_R^\dagger ) +  \right. \nn \\
&& \left.  \mbox{Tr}(\tilde{\Phi}^\dagger \Phi ) 
 \mbox{Tr}(\Delta_L \Delta_L^\dagger )  \right] 
 \, + \, \alpha_3 \left[ \mbox{Tr} (\Phi \Phi^\dagger 
\Delta_L \Delta_L^\dagger ) +\mbox{Tr} ( \Phi^\dagger \Phi 
\Delta_R \Delta_R^\dagger ) \right] + \nn \\   
 && \beta_1 \, \left[ \mbox{Tr}
 (\Phi \Delta_R \Phi^\dagger  \Delta_L^\dagger ) + 
 \mbox{Tr}(\Phi^\dagger \Delta_L  \Phi \Delta_R^\dagger ) \right]  \, + \, 
  \beta_2 \,  \left[ \mbox{Tr}
    (\tilde{\Phi} \Delta_R \Phi^\dagger  \Delta_L^\dagger ) 
 + \right. \nn \\
 && \left.   
 \mbox{Tr}(\tilde{\Phi}^\dagger \Delta_L  \Phi \Delta_R^\dagger ) \right] 
 + \beta_3 \left[
 \mbox{Tr}(\Phi \Delta_R \tilde{\Phi}^\dagger  \Delta_L^\dagger ) +  
 \mbox{Tr}(\Phi^\dagger \Delta_L  \tilde{\Phi} \Delta_R^\dagger ) \right] 
\nn \eea
where we have written out each term completely to display
the full parity symmetry.
Note that as a consequence of the discrete left-right symmetry 
all terms in the potential are self-conjugate, except for 
the $\alpha_2$ one; therefore $\alpha_2$ is the only parameter which
may be complex. The potential (\ref{pot}) is 
not invariant under the exchange of the fields 
$\phi_1^0 \leftrightarrow \phi_2^{0*}$. 
One can restore this symmetry by setting $\beta_2 = \beta_3$, 
$\alpha_3=0$ and $\alpha_2$ real. 
Then all the parameters in the scalar potential 
have to be real and it is CP conserving.
In that case, spontaneous CP violation can occur even with the 
minimal scalar sector described above \cite{bb}.
However, as explained in the introduction this model can not lead to 
successful electroweak baryogenesis and we shall not consider it here.

Since we will not discuss the CP violation aspect
of the BAU generation, in our analysis we assume CP conservation
and take $\alpha_2$ to be real. 
It can be shown \cite{despa} that, without fine tuning, the $\beta_i$ 
terms spoil the seesaw mechanism by inducing a direct Majorana mass
term for the left-handed neutrino, unless 
$|\beta_i| \le 10^{-7} - 10^{-8}$ or the right scale is very large.
As a result, in realistic left-right symmetric models 
with a low right scale ($v_R \sim$ 1 TeV) and no fine tuning,  
the effects of such terms will be 
negligible and it is a good approximation to assume that 
they vanish. Therefore, we set $\beta_i=0$ in the rest of the paper
%\footnote{If $\alpha_3 =0$, the $\beta_i$ couplings are not so strongly 
%constrained, and thus our conclusions can not be generalized 
%to that case.}. 
This choice will also avoid the unwanted presence of too large FCNC
which could enter in conflict with experimental data.

Only the neutral components of the scalar fields, 
$\phi_1^0, \phi_2^0, \delta_L^0, \delta_R^0$,
can acquire vevs without violating electric charge.
If $\Delta_L$ or $\Delta_R$ acquire a vev, then $B-L$ is 
necessarily broken, and if 
$\langle \Delta_L \rangle \neq \langle \Delta_{R} \rangle$
parity breakdown is also ensured. 
Thus the correct pattern of symmetry breaking is achieved by
\beq
\langle \Phi \rangle = \frac {1}{\sqrt{2}}  \pmatrix{k_1 & 0 \cr
0 &k_2 } \;\; ,\;\;  
\langle \Delta_{L,R} \rangle  =  \frac {1}{\sqrt{2}}
\pmatrix{ 0&0\cr v_{L,R}&0}
\label{vevs}
\eeq
where $k_1,k_2,v_L$ and $v_R$ are real, and phenomenologically the 
hierarchy $v_R \gg k_1,k_2 \gg v_L$ is required.
Moreover, when the $\beta$ parameters in the scalar potential vanish
then $v_L=0$ \cite{sm,despa}, so we neglect $v_L$ in the following.

In the tree-level scalar potential we have thus 14 free parameters, 
plus the zero temperature vevs. 
Three of these parameters can be fixed
by minimizing the zero-temperature tree-level potential, i.e. 
imposing the vanishing of the first derivatives of 
{\bf V} at $(k_1,k_2,v_R)$, which leads to the 
relations 
\bea
\mu_1^2 &=& \frac{\alpha_1}{2} v_R^2 - \frac{\alpha_3}{2} \frac{k_2^2 v_R^2}{
(k_1^2 -k_2^2)} + \lambda_1 (k_1^2 + k_2^2) + 2 \lambda_4 k_1 k_2
\nonumber \\
\mu_2^2 &=&\frac{\alpha_2}{2} v_R^2 + \frac{\alpha_3}{4} \frac{k_1 k_2 v_R^2}{
(k_1^2 -k_2^2)} +  ( 2 \lambda_2 + \lambda_3) k_1 k_2 + \frac{\lambda_4}{2}
(k_1^2 + k_2^2) 
\label{min} 
\\
\mu_3^2 &=& \frac {\alpha_1}{2}(k_1^2 + k_2^2) + \rho_1 v_R^2 
+2 \alpha_2 k_1 k_2 + \frac {\alpha_3}{2} k_2^2
\nonumber
\eea

Before writing down the finite temperature effective potential, let us
discuss briefly the values that can be taken by the $\alpha_i \;(i=1,2,3)$
parameters. From the minimization conditions (\ref{min}), one can see 
that to obtain $\mu_1$ and $\mu_2$ of order of the weak scale
(and hence phenomenologically acceptable values of 
$k \equiv \sqrt{k_1^2 + k_2^2}$) 
the $\alpha_i$ should be of order ${\cal O} (k^2/v_R^2) \ll 1$. 
Otherwise, $\mu_1, \mu_2$ 
would naturally be of order of the right scale, $v_R$.
But this is by no means an artificial fine tuning, as the $\alpha_i$
parameters govern the doublet-triplet mixing and therefore we do expect
them to be of that order.
We shall take advantage of  this fact to obtain an approximate analytic 
expression for the finite temperature effective potential in the next section.

The masses for the relevant degrees of freedom of the theory
in the background of the fields $k_1,k_2,v_R$
are given in the appendices.

\section{Finite Temperature Effective Potential}

The main tool for the study of the electroweak phase transition in the 
left-right symmetric model described above is the one-loop, 
daisy improved finite-temperature effective potential of the model.
We are actually interested in the dependence of the potential on 
$k_1= {\rm Re} \; \phi_1^0/\sqrt{2}, k_2= {\rm Re} \;\phi_2^0/\sqrt{2}$
and $v_R = {\rm Re} \; \delta_R^0 /\sqrt{2}$. 
It can be readily computed by the usual methods \cite{dj} and is 
given by 
\beq
V_{eff}(k_i,v_R,T) = V(k_i,v_R) + V_1(k_i,v_R,T)
+V_{daisy}(k_i,v_R,T) 
\label{pot1}
\eeq
where $V(k_i,v_R)$ is the tree-level potential (\ref{pot}), 
\bea 
V_1(k_i,v_R,T) &=& \frac {T^4}{2 \pi^2}
\sum_i n_i J_i \left[ \frac{m_i^2(k_i,v_R)}{T^2} \right] \;, 
\\
V_{daisy}(k_i,v_R,T) &=& - \frac {T}{12 \pi}
\sum_i n_i 
\left[ \overline{m}_i^3(k_i,v_R,T) - m_i^3(k_i,v_R) \right] \ .  
\eea
The sum runs over all the particles in the model, $n_i$ is 
the corresponding number of degrees of freedom,
taken negative for fermions, and 
$m_i^2(k_i,v_R)$ is the tree-level mass of the particle $i$ 
in presence of the background fields $k_1,k_2,v_R$.
The functions $J_i = J_+ (J_-)$ for bosons (fermions) are
given by
\beq
J_{\pm}(y^2) = \int_0^\infty dx x^2 \; \log 
\left(1 \mp e^{- \sqrt{x^2+y^2}} \right) 
\eeq
The last term in (\ref{pot1}), $V_{daisy}(k_i,v_R,T)$, 
is a correction coming 
from the resummation of the leading infrared divergent higher-loop 
contributions, associated with the so-called daisy diagrams. The sum 
runs over bosons only.   
The masses $\overline{m}_i^2(k_i,v_R,T)$ are obtained from the 
$m_i^2(k_i,v_R)$ by adding the leading $T$-dependent self-energy 
contributions, which are proportional to $T^2$.
In the contribution of the longitudinal gauge boson degrees of freedom, 
there is a suppression due to the temperature dependent Debye mass.
A simple treatment is just to drop the longitudinal 
contribution \cite{dine}, and we follow this prescription. 

For values of the fields such that 
$m_i(k_i,v_R)/T < 1$, we can expand $J_{\pm}$ as \cite{dj}
\footnote{In fact, it may be shown numerically that the $m/T$
expansion is a good approximation up to $m/T \sim 2.2 (1.6)$ for
bosons (fermions) \cite{ah}.}
\bea
J_+ (m^2/T^2)&=& -\frac{\pi^4}{45} + \frac{\pi^2 m^2}{12 T^2}
-\frac{\pi}{6} \left(\frac{m^2}{T^2}\right)^{3/2}
+ {\cal O} \left(\frac{m^4}{T^4} \log \frac m T \right)
\\
J_-(m^2/T^2) &=& \frac{7 \pi^4}{360} - \frac{\pi^2 m^2}{24 T^2}
+ {\cal O} \left(\frac{m^4}{T^4} \log \frac m T \right)
\eea
In left-right symmetric models one expects two phase transitions
\cite{cv}:
one at $T=T_R={\cal O}(v_R) \sim 1$ TeV \cite{lin}, 
where $SU(2)_R$ is spontaneously broken, and the other at  
$T=T_L={\cal O}(k) \sim 250$ GeV.
Hence at temperatures much higher than $T_R$ down to $T=T_R$, 
$v_R = k = 0$ will be 
the minimum of the effective potential (\ref{pot1}).
At $T=T_R$, two degenerate minima exist: one for 
$v_R = k = 0$ and a new one at $k=0$, $v_R = v_R(T_R)$.
For temperatures $T < T_R$, the right triplet field $v_R$ will
settle down near the minimum given by $v_R = v_R(T_R)$, 
which will slowly evolve to its zero temperature value.
In our analysis we focus on the left-sector phase transition, 
and we just assume that by the time it occurs, equilibrium has again 
been attained after the right phase transition, so that
we can reliably use the finite temperature effective potential 
(\ref{pot1}).

Near the electroweak phase transition temperature, $T_L$,
the high-temperature expansion of $J_\pm$ is not valid 
for particles with mass of order $v_R \gg T_L$.
The contribution due to these particles is Boltzmann suppressed, 
and in the limit $m_i(k_i,v_R) \gg T$ it 
reduces to \cite{ah}
\beq
V_1(k_i,v_R,T) \sim 
\sum_i \frac {n_i T^2}{(2 \pi)^{3/2}} m_i^2
\sqrt{\frac {T} {m_i}} e^{-m_i/T} 
\left[ 1 + \frac{15 T}{8 m_i} + 
O \left( \frac{T^2}{m_i^2} \right) \right] \ .  
\eeq 
The exponential factor in the previous expression allows us,
within good approximation, to neglect the effect of particles 
with masses of the $v_R$ scale. 
We then have to identify the heavy ${\cal O}(v_R)$ 
degrees of freedom, by diagonalizing the mass matrices given in the
appendices,
and remove their contribution from eq. (\ref{pot1}).

In the case of the gauge bosons, we see from eqs. 
(\ref{charged}), (\ref{neutral}) that only the 
$W_1$ and $Z_1$ bosons should be included in eq. (\ref{pot1}), 
since $W_2$ and $Z_2$ get masses of order $v_R$. 
In the limit $k^2 \ll v_R^2$, the $W_1$ mass is just given by the 
(11) entry of the mass matrix (\ref{ma1}), while the $Z_1$ mass 
can be approximated by
\beq
M^2_{Z_1} \simeq 
\frac{g^2(g^2 + 2 g'^2)}{4(g^2 + g'^2)} (k_1^2 + k_2^2).
\eeq
Since the electric charge formula, eq.(\ref{elc}), implies that
\beq
\frac{1}{e^2} = \frac{2}{g^2} + \frac{1}{g'^2},
\eeq
a  Standard Model type  relation  for the light gauge bosons  
is also valid in the left-right symmetric model,
$M^2_{Z_1} = M^2_{W_1} /
\cos^2\theta_W $, with 
$cos^2\theta_W = (g^2 + g'^2)/(g^2 + 2 g'^2)$.

With respect to the fermions, only the right-handed neutrinos get 
masses of order $v_R$.
The contribution to the effective potential 
of quarks and charged leptons is proportional to the Yukawa 
couplings $(h,\tilde{h})$, 
so we neglect all of them except for the third 
generation of quarks. 
The effect of the light neutrinos is even more suppressed, 
since their masses are of order $k^2/v_R$ due to the see-saw
mechanism.

Let us finally discuss the scalar sector spectrum.
Both doubly charged scalar fields, $\delta_R^{++}, \delta_L^{++}$,  
acquire masses of order $v_R$ and hence decouple. 
In the singly charged Higgs sector, $\delta_L^{+}$ is an 
eigenstate with mass of ${\cal O}(v_R)$, while 
$\phi_1^+ , \phi_2^+ , \delta_R^{+}$ mix among themselves.  
However, the mass matrix elements which relate the 
triplet with the doublet components are of the type
$\alpha_i k v_R$, which are negligible 
with respect to the terms ${\cal O}(\lambda_i k^2)$,
since $\alpha_i = {\cal O}(k^2/v_R^2)$
and $\lambda_i = {\cal O}(1)$.
Thus, neglecting the doublet-triplet mixing, 
$\delta_R^{+}$ can be identified with the pseudo-Goldstone 
boson eaten by the $W_R$, which in a general $R_\xi$ gauge 
will not contribute to the effective potential (\ref{pot1})
near $T_L$.
The two remaining charged scalar fields get electroweak scale 
masses, which can be calculated analytically by diagonalizing 
the corresponding $2 \times 2$ submatrix, given in 
appendix B.

Since we have assumed that the tree-level scalar potential is 
CP conserving, in the neutral Higgs sector scalars and 
pseudo-scalars do not mix, and we are left with two $4 \times 4$ mass 
matrices. From the one corresponding to the imaginary parts of 
the neutral fields, we see that neither $\delta_R^i$ nor 
$\delta_L^{i}$ contribute to $V_{eff}(k_i,v_R,T)$; 
the former is the pseudo-Goldstone boson
eaten by the $Z'$, while the latter acquires a right scale mass.
Concerning the real parts, $\delta_L^r$ decouples because it is also 
a heavy eigenstate, and neglecting terms of the type 
$\alpha_i k v_R$ so does $\delta_R^r$.
The field dependent masses of the light eigenstates can be found in 
appendix B.

Then, the field-dependent part of the finite temperature 
effective potential near the electroweak phase transition may be 
approximated by:

\bea
V_{eff}(k_i,v_R,T) &=& V(k_i,v_R) + 
\frac {T^2}{24} \left\{ 2 (2 \alpha_1 + \alpha_3) v_R^2 +
\left[ 24 \lambda_4 + 12 h \tilde{h} \right] k_1 k_2 \right.
\label{pot2}
\\
& & + \left. \left[ 
10 \lambda_1 + 4 \lambda_3 + \frac{3}{2} g^2 + \frac{3}{4} \frac{g^2
(g^2 + 2 g'^2 )}{(g^2 + g'^2)} +
3 (h^2 + \tilde{h}^2) \right] (k_1^2 + k_2^2)  
\right \}
\nonumber \\
%V_{daisy}(k_i,v_R,T) &=& 
&-&\frac {T}{12 \pi} \left\{ 
\sum_{j=1}^8 [\overline{m}_j^2(k_i,v_R,T)]^{3/2}
+ 4 \left[ \frac{g^2}{4} (k_1^2 + k_2^2) \right]^{3/2}
+ 2 \left[ \frac{g^2
(g^2 + 2 g'^2 )}{4(g^2 + g'^2)} (k_1^2 + k_2^2) \right]^{3/2} \right\}
\nonumber
\eea
where the sum runs only over the bidoublet scalar degrees of freedom.

Notice that, within reasonable approximations, we have found that 
in the scalar sector only the bidoublet $\Phi$ can give a sizeable 
contribution to the finite temperature effective potential near 
$T_L$, together with the fermions and SM gauge bosons.
The effective theory at temperatures of order $T_L$ contains then 
the same degrees of freedom as a two Higgs doublet model.
However, a careful look at the part of the tree-level potential 
involving the bidoublet shows that
some of the scalar couplings of the most general two Higgs model 
in the left-right symmetric model are constrained or vanish, 
while the coupling $\lambda_4$, usually taken to be zero in two 
Higgs models, is present in our case. 
Therefore, although the electroweak phase transition in two Higgs 
doublet models has been extensively studied in the literature \cite{2h}, 
the results can not be extrapolated to the left-right symmetric model 
in a straight forward way, and it is worth to perform a new analysis
within the (different) parameter space relevant for this case.

\section{Numerical results}

We shall now use the effective potential (\ref{pot2}) to calculate the 
critical temperature and the location of the minimum at the 
critical temperature. 
We define the critical temperature $T_c$ as the 
value of $T$ at which the determinant of the second 
derivatives of $V_{eff}(k_i,v_R,T)$ at $k=0$ vanishes:
\beq
{\rm det} \left [ 
\frac {\partial^2 V_{eff}(k_i,v_R,T_c)}
{\partial k_i \partial k_j} \right]_{k=0} = 0 \ . 
\label{tc}
\eeq
In fact, the phase transition starts at $T = T_D$, where
$T_D$ is the temperature at which there are two degenerate 
minima, by tunnelling. 
At $T_c$ there is no longer any barrier in some direction between 
what was the minimum at the origin and the new minimum away from 
the origin, and condensation of the scalar fields can progress
rapidly without any suppression from a tunnelling factor. 

%In fact, the actual temperature $T_0$ at which the phase transition 
%occurs satisfies the inequalities $T_c < T_0 < T_D$, 

The effective potential $V_{eff}(k_i,v_R,T_c)$ is a 
function of the three temperature-dependent vevs, 
$k_1(T), k_2(T), v_R(T)$; however we expect 
$v_R(T_c) \simeq v_R(T=0)$ and therefore we approximate 
$v_R(T_c)$ by its zero temperature value. 
Within this approximation, 
we solve numerically eq. (\ref{tc}), and once $T_c$ is determined we 
minimize (numerically) the potential $V_{eff}(k_i,v_R,T_c)$ 
and find the minimum $[k_1(T_c),k_2(T_c)]$. 
Then we compute the quantity of interest concerning the strength of the 
electroweak phase transition, that is, the ratio $k(T_c)/T_c$
where $k(T_c) \equiv \sqrt{k_1^2(T_c) + k_2^2(T_c)}$.

Our procedure is the following: we use the minimization conditions 
at zero temperature (\ref{min}) to fix three of the unknown 
parameters in the effective potential (\ref{pot2}).
The experimental constraint on the weak scale
\beq
k^2 \equiv k_1^2 + k_2^2 = (246 {\rm GeV})^2
\label{wscale}
\eeq
eliminates one more, while the Yukawa couplings 
$h, \tilde{h}$
can be determined from the masses of the third generation of quarks, 
for which we take $m_{top} = 175$ GeV and $m_b = 4.5$ GeV.

The finite temperature effective potential (\ref{pot2}) still 
depends on a large number of free parameters in the scalar sector. 
However, only some of them are relevant to determine the 
critical temperature and the position of the minimum, namely
$\lambda_1,\lambda_2,\lambda_3, \lambda_4$, $k_1(T=0)$
and $\alpha_3$, which only appears in the combination 
$\alpha_3 v_R^2$. We generate randomly values of these
parameters in the ranges $|\lambda_i| \leq 1/2$ 
(so that the use of the perturbative effective potential is reasonable),
$|k_1(T=0)| \leq 246$ GeV 
and $|\alpha_3 v_R^2| \leq (246 {\rm GeV})^2/10$, 
which takes into account that 
$\alpha_i \sim {\cal O} (k^2/v_R^2)$ for realistic left-right 
symmetric models.

There are further restrictions on this parameter space, 
due to zero temperature requirements. 
First, the potential must be bounded from below, which leads to 
the set of constraints:
\beq
\lambda_1 > 0 \;\;\;\;
\lambda_1 - |\lambda_4| > 0  \;\;\;\;
2 \lambda_2 + \lambda_3 - \lambda_1 > 0
\label{c1}
\eeq

Finally,  
in order to obtain the correct symmetry breaking pattern at zero 
temperature, we also require that the scalar vevs do not break 
electromagnetism and that the squared masses of fluctuations 
about these vevs are positive. That is, the eigenvalues of the 
light scalar mass matrices at zero temperature (see appendix B)
should be positive, once the relations (\ref{min}) have been 
used.
For any random set ($\lambda_1,\lambda_2,\lambda_3, \lambda_4, 
k_1(T=0), \alpha_3 v_R^2$) satisfying the above conditions, 
we calculate the critical temperature $T_c$ according to the definition 
(\ref{tc}), minimize the effective potential at $T_c$ with respect 
to $k_1(T_c),k_2(T_c)$, and obtain $k(T_c)/T_c$.

Since we use the high temperature expansion of the one-loop 
effective potential, we need to verify that such approximation is
valid at $T_c$. Thus, once the vevs $k_i(T_c)$ have been determined, 
we compute the value of all the masses which enter in the effective 
potential (\ref{pot2}) and impose the condition 
\beq
\frac{m(T_c)}{T_c} < 1.6
\eeq
For the sets of parameters excluded by this condition, the
high temperature expansion used would be questionable. 

In Fig. 1 we plot the ratio $k(T_c)/T_c$ against the lightest 
scalar mass $m_1$, corresponding to a sample of 500 points in the 
parameter space which passed our selection criterion. 
As we see, there is a sizeable fraction which satisfies the
condition for preserving the baryon asymmetry,
$k(T_c)/T_c > 1$, and corresponds to experimentally allowed 
values of the lightest scalar mass, $m_1 >  50$ GeV \cite{fin}. 
We find this result to be particularly interesting, given the
relatively large number of potential signatures of such a 
model in future experiments \cite{phen} and the small number
of free parameters to adjust the remaining phenomenology \cite{pil}.
 
In Figs. 2, 3, 4 and 5 we show the frequency of occurrence of the 
(zero temperature) masses corresponding to the light physical 
scalars, in the allowed baryon preserving region for a sample
of 9000 points. 
They range from about 50 GeV to 250 GeV.
The masses of the lightest neutral scalar and the charged ones 
are peaked about 110 GeV, while the pseudo-scalar 
and heavy neutral scalar mass 
distributions are broader and centered in a somehow higher value 
$\sim 150$ GeV. 
So we conclude that there is no significant contradiction 
with experimental bounds in the baryon preserving cases 
found.

\section{Conclusions}

We have analyzed the electroweak phase transition in left-right 
symmetric models with a scalar sector consisting of 
a bidoublet and two triplets. Within reasonable simplifying 
assumptions about the scalar couplings, we find 
regions of parameter space which are consistent with the present 
experimental bound on the Higgs mass and with a sufficiently strong 
first-order electroweak phase transition, eq. (\ref{sph}).
We have also obtained the scalar spectrum for these 
phenomenologically acceptable values of the parameters.

In this paper we have focused on the requirement that the sphaleron 
processes be sufficiently suppressed after the electroweak phase 
transition, to preserve the produced baryon asymmetry.
Once we have shown that the transition can be strongly enough
first order, a detailed calculation of the baryon asymmetry
generated during the electroweak phase transition
in the framework of left-right symmetric models would be very
interesting. As mentioned in the introduction, there are 
estimates of this quantity in the literature \cite{jm}-\cite{rt}
but they do not include some relevant effects and lead to
different results.
In principle, the baryon asymmetry in the class of left-right 
symmetric models considered here will 
be generated in much the same manner as in two Higgs doublet 
models, where it has been computed by several groups 
\cite{nkc}-\cite{cpr}. Some of 
these calculations seem to indicate that enough 
baryon asymmetry can be generated, so we expect that 
this will also be the case in left-right symmetric models.

\section*{Acknowledgements}

We thank D. Comelli, J. Moreno, M. Pietroni, M. Quir\'os 
and G. Senjanovic for useful discussions,
and E. Roulet for comments on the manuscript. 
This work was supported in part by the CICYT contract AEN-96/1718,
the DGICYT contract PB95-1077 and by EEC under the 
TMR contract ERBFMRX-CT96-0090.
One of us (G.B.) acknowledges a post-doctoral fellow
of the Teilchen und Mittelrnergiephysik Graguiertenkolleg
of the University of Mainz.

\vspace{1cm}

\vskip 2.truecm

\setcounter{section}{0}
\def\theequation{\Alph{section}.\arabic{equation}}
\begin{appendix}
\setcounter{equation}{0}
\section{Gauge boson eigenstates}
For the sake of completeness in this appendix we derive physical gauge
boson eigenstates (eigenstates of mass matrices) and their 
eigenvalues. Remember that our scalar sector consists of one bidoublet
and one set of left-right symmetric lepton-number carrying triplets
with the pattern of symmetry breaking given in eq. (\ref{vevs}).

The piece of the Lagrangian containing their covariant derivatives is
\bea  
{\cal L}_{D}=\mbox{Tr}(D_\mu\Delta_{_L})^\dagger(D^\mu\Delta_{_L})
  +\mbox{Tr}(D_\mu\Delta_{_R})^\dagger(D^\mu\Delta_{_R})+
\mbox{Tr} (D_\mu\Phi)^\dagger(D^\mu\Phi)
\label{cd} 
\eea
where
\bea   D_\mu\Delta_{_L}&=&\partial_\mu\Delta_{_L}
      +\frac{1}{2}ig \left[ \vec{\tau}\cdot\vec{W}_L\Delta_{_L}
        - \Delta_{_L} \vec{\tau} \cdot\vec{W}_L \right]
       +\frac{1}{2}ig'B\Delta_{_L},\nn \\
      D_\mu\Delta_{_R}&=&\partial_\mu\Delta_{_R}
      +\frac{1}{2}ig\left[ \vec{\tau}\cdot\vec{W}_R\Delta_{_R}
        -\Delta_{_R} \vec{\tau} \cdot\vec{W}_R \right]
      +\frac{1}{2}ig'B\Delta_{_R}, \nn \\
      D_\mu\Phi&=&\partial_\mu\Phi
      +\frac{1}{2}ig(\vec{\tau}\cdot\vec{W}_L\Phi
      -\Phi\vec{\tau}\cdot\vec{W_R})        
\label{cd2}
\eea
Then in this model there are seven gauge bosons: 
 four charged ones, the $W_{L,R}^1$ and  $W_{L,R}^2$ and
three neutral ones, $W_{L,R}^3$ and $B$.
When the Higgs multiplets acquire their vevs (see eq. (\ref{vevs})) 
the interaction bosons get their masses.

By inspecting the Lagrangian, it is easy to see that the mass terms
for the charged bosons are
\bea 
 {\cal L}_{mass}^c=(\begin{array}{cc}W_L^+&W_R^+\end{array})M^{c}
     \left(\begin{array}{c}W_L^-\\W_R^-\end{array}\right)  
 \eea
where $W^{\pm}$ are defined by
$ W^{\pm}=\frac{1}{\sqrt{2}}(W^1\mp W^2)   $
and $M^{c}$ is
\bea 
 M^{c}=\frac{g^2}{4}\left(\begin{array}{cc}
    k_1^2+k_2^2&-2k_1 k_2
\\-2k_1 k_2&v_R^2+k_1^2+k_2^2\end{array}\right)  
\label{ma1}
\eea
While that of the neutral sector has the form 
\bea 
 {\cal L}_{mass}^{n}=
     \frac{1}{2}(\begin{array}{ccc}W_L^3&W_R^3&B\end{array})M^n
     \left(\begin{array}{c}W_L^3\\W_R^3\\B\end{array}\right) 
\eea
where the $M^{n}$ is given by
\bea  M^{n}=\frac{1}{4}\left(\begin{array}{ccc}
     g^2(k_1^2+k_2^2)&  -g^2(k_1^2+k_2^2)   &   0\\
       -g^2(k_1^2+k_2^2)   &g^2(v_R^2+k_1^2+k_2^2)&   -gg'v_R^2\\
         0       &     -gg'v_R^2     & g'^2 v_R^2
     \end{array}\right)
\label{ma2}
\eea
The diagonalization of (\ref{ma1}) and (\ref{ma2}) gives the masses
of the charged $W_{1,2}^\pm$ and neutral $A$ and $Z_{1,2}$ physical 
fields, they are 
\bea 
 M_{W_{1,2}}^2=\frac{g^2}{8}
  \left[v_R^2+2(k_1^2+k_2 ^2)\mp
\sqrt{v_R^4+16(k_1 k_2)^2}\right]
\label{charged} 
\eea
\bea
 M_{Z_{1,2}}^2=C\mp\sqrt{C^2-4D}  
\label{neutral}
\eea
with
\[  C=\frac{1}{8}[(g^2+g'^2)v_R^2+2g^2(k_1^2+k_2^2)]  \]
\[  D=\frac{1}{64}g^2(g^2+2g'^2)(k_1^2+k_2^2)v_R^2   \]
and
\bea
M_A = 0
\eea

\section{Higgs masses}
\setcounter{equation}{0}
Here we give a variety of useful result for the mass-squared matrices of the
various Higgs sectors before the first derivative constraints have been
substituted.
The mass matrices are symmetric.

\subsection{Neutral scalar mass matrix}
We first compute the mass matrix corresponding to the real components 
of the neutral scalar fields in the 
$ \{\phi_1^r \, , \, \phi_2^r \, , \, \delta_R^{r} \, , \, \delta_L^{r}\}$
basis.

\bea
{\cal{M}}^{\mbox{\tiny{Re}}^2}_{11} & = &
 -\mu_1^2 + \lambda_1 ( 3 k_1^2 + k_2^2 ) +
4 \lambda_2 k_2^2 + 2 \lambda_3 k_2^2  + 6 \lambda_4 k_1 k_2 +  \frac{1}{2}
 \alpha_1 v_R^2 
 \nn \\
{\cal{M}}^{\mbox{\tiny{Re}}^2}_{12} &= & -2 \mu_2^2 + 
 k_1 k_2  \left( 2 \lambda_1 + 8 \lambda_2 + 4 \lambda_3 \right)
 + 3 \lambda_4 ( k_2^2 + k_1^2 )+  \alpha_2  v_R^2  
 \nn\\
{\cal{M}}^{\mbox{\tiny{Re}}^2}_{13} & =& \alpha_1 k_1 v_R + 2 \alpha_2 k_2 v_R 
\nn \\
{\cal{M}}^{\mbox{\tiny{Re}}^2}_{14}&=& 0
\nn \\
{\cal{M}}^{\mbox{\tiny{Re}}^2}_{22}&=& - \mu_1^2 + \lambda_1 \left( 3 k_2^2 + 
k_1^2 \right) + 2 k_1^2 ( 
2 \lambda_2 +\lambda_3 ) + 6 \lambda_4 k_1 k_2  
 + \frac{1}{2} (\alpha_1 + \alpha_3) v_R^2 
\nn \\
{\cal{M}}^{\mbox{\tiny{Re}}^2}_{23}&=&2  \alpha_2 k_1 v_R +  \alpha_1 k_2 v_R
+ \alpha_3 k_2 v_R
\nn \\
{\cal{M}}^{\mbox{\tiny{Re}}^2}_{24} &=& 0
\nn
\\
{\cal{M}}^{\mbox{\tiny{Re}}^2}_{33} 
&=& - \mu_3^2 + 3 \rho_1 v_R^2  
  + 2 \alpha_2 k_1 k_2 + \frac{1}{2} \alpha_1
 (k_1^2 + k_2^2) + \frac{1}{2}\alpha_3 k_2^2
\nn \\
{\cal{M}}^{\mbox{\tiny{Re}}^2}_{34}
&=& 0
\nn \\
{\cal{M}}^{\mbox{\tiny{Re}}^2}_{44} &=& -\mu_3^2 + 
 \frac{1}{2} \rho_3 v_R^2 + \frac{1}{2}
\alpha_1 (k_1^2 + k_2^2) + 2 \alpha_2 k_1 k_2 + \frac{1}{2}  \alpha_3 k_2^2
\eea

As explained in section 3, near the electroweak phase transition temperature
only the light states contribute to the effective potential in 
eq. (\ref{pot1}) and are relevant for our analysis.
Within the approximation $k^2 \ll v_R^2$, those are the bidoublet components 
$\phi_1^r \, , \, \phi_2^r$, and the corresponding  mass matrix is just the 
$2 \times 2$ submatrix obtained from the entries (11),(12) and (22) above.

\subsection{Neutral pseudo-scalar mass matrix}
In a manner similar to the previous section, we compute the 
mass matrix corresponding to the imaginary components 
of the neutral scalars, in the basis
$\{\phi_1^i \, , \, \phi_2^i \, , \, \delta_R^{i} \, , \, \delta_L^{i}\}$.

\bea
{\cal{M}}^{\mbox{\tiny{Im}}^2}_{11} & = & -\mu_1^2 + \lambda_1 ( k_1^2 + k_2^2 ) -
4 \lambda_2 k_2^2 + 2  \lambda_3 k_2^2  + 2 \lambda_4 k_1 k_2 +  \frac{1}{2}
 \alpha_1 v_R^2 
 \nn \\
{\cal{M}}^{\mbox{\tiny{Im}}^2}_{12} &= & 2 \mu_2^2 -  8 \lambda_2 k_1 k_2 
 -  \lambda_4 ( k_2^2 + k_1^2 ) -  \alpha_2  v_R^2  
 \nn\\
{\cal{M}}^{\mbox{\tiny{Im}}^2}_{13} & =& 0
\nn \\
{\cal{M}}^{\mbox{\tiny{Im}}^2}_{14}&=& 0
\nn \\
{\cal{M}}^{\mbox{\tiny{Im}}^2}_{22}&=& - \mu_1^2 + \lambda_1 \left(  k_2^2 + 
k_1^2 \right) + 2 k_1^2 ( 
- 2 \lambda_2 +\lambda_3 ) + 2 \lambda_4 k_1 k_2  
 + \frac{1}{2} (\alpha_1 + \alpha_3) v_R^2 
\nn \\
{\cal{M}}^{\mbox{\tiny{Im}}^2 }_{23}&=&0
\nn \\
{\cal{M}}^{\mbox{\tiny{Im}}^2}_{24} &=& 0
\nn
\\
{\cal{M}}^{\mbox{\tiny{Im}}^2}_{33} 
&=& - \mu_3^2 + \rho_1 v_R^2  
  + 2 \alpha_2 k_1 k_2 + \frac{1}{2} \alpha_1
 (k_1^2 + k_2^2) + \frac{1}{2} \alpha_3 k_2^2
\nn \\
{\cal{M}}^{\mbox{\tiny{Im}}^2}_{34} 
&=& 0
\nn \\
{\cal{M}}^{\mbox{\tiny{Im}}^2}_{44} &=& -\mu_3^2 + 
 \frac{1}{2} \rho_3 v_R^2 + \frac{1}{2}
\alpha_1 (k_1^2 + k_2^2) + 2 \alpha_2 k_1 k_2 + \frac{1}{2}  \alpha_3 k_2^2
\eea

Again, in the limit $k^2 \ll v_R^2$, the light states are the bidoublet 
components $\phi_1^i \, , \, \phi_2^i$, and their mass matrix is given 
by the entries (11), (12) and (22) of 
${\cal{M}}^{\mbox{\tiny{Im}}^2}$.

\subsection{Singly charged Higgs mass matrix}
The singly charged Higgs mass matrix, in the
$ \{\phi_1^+ \, , \, \phi_2^+ \, , \, \delta_R^{+} \, , \, \delta_L^{+} \}$ 
basis, is 
 
\bea
{\cal{M}}_{11}^{+ 2}  & = &  -\mu_1^2 + \lambda_1 (k_1^2 + k_2^2 ) + 
2 \lambda_4 k_1 k_2 + \frac{1}{2} ( \alpha_1+ \alpha_3 ) v_R^2 
\nn \\
{\cal{M}}_{12}^{+ 2} & = & -\alpha_2 v_R^2 + 2 \mu_2^2 - \lambda_4 (k_1^2
+ k_2^2 ) - 2 k_1 k_2  \left( \lambda_3 + 2 \lambda_2 \right)
\nn \\
{\cal{M}}_{13}^{+ 2} & = &  \frac{1}{2\sqrt{2}} \alpha_3 k_1 v_R \nn \\
{\cal{M}}_{14}^{+ 2} & = & 0
\nn \\
{\cal{M}}_{22}^{+ 2} & = & - \mu_1^2 + \frac{1}{2} \alpha_1 v_R^2 + \lambda_1
(k_1^2 + k_2^2) + 2 \lambda_4 k_1 k_2 
\nn \\
{\cal{M}}_{23}^{+ 2} & = & \frac{1}{2\sqrt{2}} \alpha_3 k_2 v_R 
\nn \\
{\cal{M}}_{24}^{+ 2} & = & 0
\nn \\
{\cal{M}}_{33}^{+ 2} & = & -\mu_3^2 + \frac{1}{2} \left( \alpha_1 
+     \alpha_3 \right) (k_1^2 + k_2^2 )+ 
2 \alpha_2 k_1 k_2 + \rho_1 v_R^2 
\nn \\
{\cal{M}}_{34}^{+ 2} & = & 0
\nn \\ 
{\cal{M}}_{44}^{+ 2} & = & -\mu_3^2 +\frac{1}{2} \left( \alpha_1 
+   \alpha_3 \right)  (k_1^2 + k_2^2 ) + 2 \alpha_2
k_1 k_2  +  \frac{1}{2} \rho_3 v_R^2 
\eea

In the limit $k^2 \ll v_R^2$, the light mass eigenstates coincide with 
$\phi_1^+ \, , \, \phi_2^+$, and their mass matrix is given 
by the entries (11), (12) and (22) of 
${\cal{M}}^{+^2}$.

\subsection{Doubly charged Higgs mass matrix} 
We now present the doubly charged Higgs mass matrix components in
the  $ \{\delta_R^{++} \, , \, \delta_L^{++} \}$ basis .

\bea
{\cal{M}}_{11}^{++ 2} & = & -\mu_3^2 + \frac{1}{2} \alpha_1  ( k_1^2 + k_2^2 )
 + 2 \alpha_2 k_1 k_2 + \rho_1 v_R^2 + 2 \rho_2 v_R^2 + \frac{1}{2} \alpha_3
k_1^2
\nn \\
{\cal{M}}_{12}^{++ 2} & = & 0
\nn \\
{\cal{M}}_{22}^{++ 2} & = & -\mu_3^2 + \frac{1}{2} \alpha_1  ( k_1^2 + k_2^2 )
 + 2 \alpha_2 k_1 k_2
 + \frac{1}{2} \rho_3 v_R^2 + \frac{1}{2} \alpha_3
k_1^2
\eea

\end{appendix}

\newpage

{\large\bf Figure captions}

\vspace{0.5cm}

\begin{itemize}

\item[{\bf Fig. 1.}] $v(T_c)/T_c$ ratio vs the ligthest scalar mass (in Gev).
\\
\item[{\bf Fig. 2.}]
Frequency distribution of the lightest neutral scalar mass (in Gev) 
for the baryon preserving cases.
\\
\item[{\bf Fig. 3.}]
Frequency distribution of the charged scalar mass (in GeV) for the baryon 
preserving cases.
\\\item[{\bf Fig. 4.}]
Frequency distribution of the pseudoscalar mass (in Gev) for the baryon 
preserving cases.
\\
\item[{\bf Fig. 5.}]
Frequency distribution of the heavy neutral scalar mass (in Gev) for the 
baryon preserving cases.

\end{itemize}

\pagebreak

\begin{figure}
\begin{center}
\epsfxsize = 13cm
\epsffile{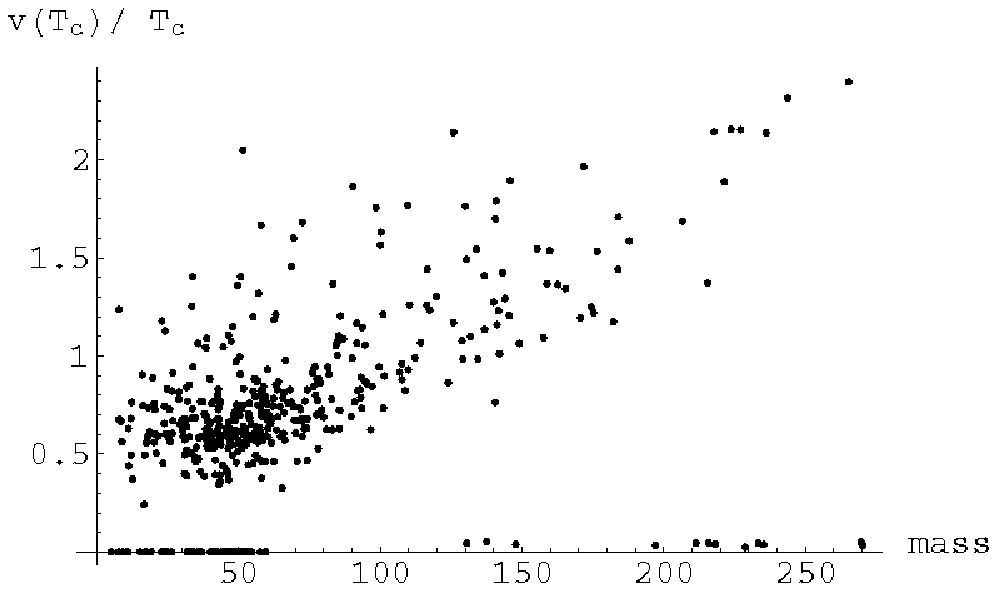}
\leavevmode
\end{center}
\end{figure}

\pagebreak

\begin{figure}
\begin{center}
\epsfxsize = 13cm
\epsffile{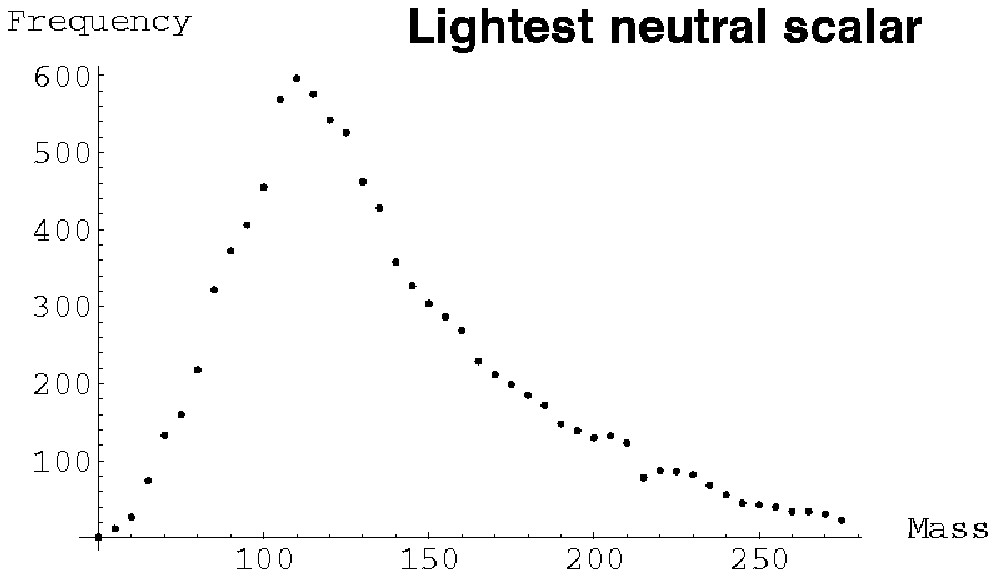}
\leavevmode
\end{center}
\end{figure}

\pagebreak

\begin{figure}
\begin{center}
\epsfxsize = 13cm
\epsffile{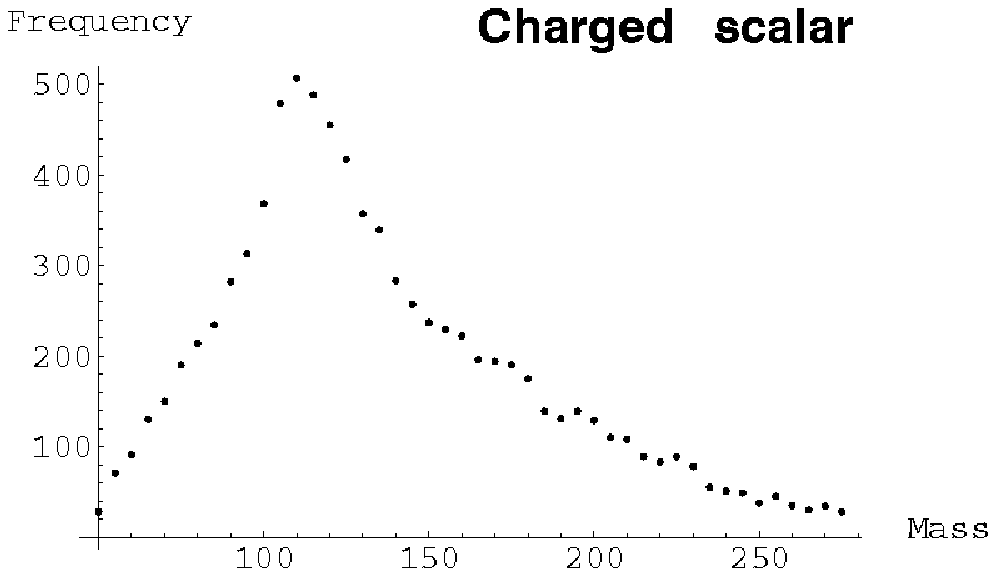}
\leavevmode
\end{center}
\end{figure}

\pagebreak

\begin{figure}
\begin{center}
\epsfxsize = 13cm
\epsffile{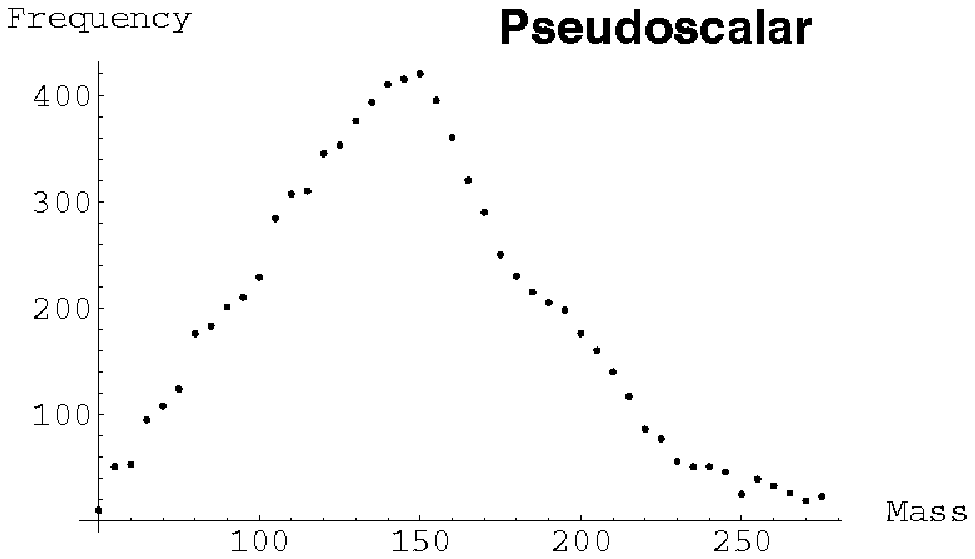}
\leavevmode
\end{center}
\end{figure}

\pagebreak

\begin{figure}
\begin{center}
\epsfxsize = 13cm
\epsffile{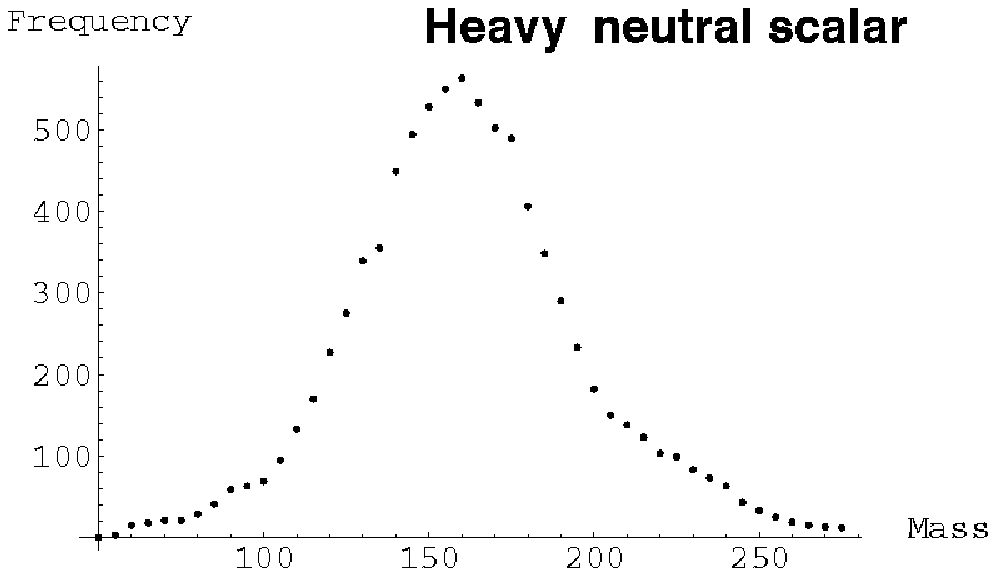}
\leavevmode
\end{center}
\end{figure}

\end{document}